\documentclass[12pt]{article}
\usepackage{sbc-template}
\usepackage{graphicx,url}
\usepackage{verbatim}
\usepackage{acronym}
\usepackage{subfigure}
\usepackage{enumitem}
\usepackage{booktabs}
\usepackage{amssymb}
\usepackage{multirow}
\usepackage{graphicx}
\usepackage{gensymb}
\usepackage{caption}
\usepackage[utf8]{inputenc}
\usepackage[brazil]{babel}
\usepackage{multirow}
\usepackage{hyperref}
\usepackage{graphicx}
\usepackage{url}
\usepackage{epsfig}
\usepackage{pifont}
\usepackage{xspace}
\usepackage{textcomp}
\usepackage[brazilian]{algxpar}
\usepackage{amsmath}
\usepackage[boxruled,linesnumbered,portuguese]{algorithm2e}

\acrodef{ITS}{\textit{Intelligent Transportation Systems}}
\acrodef{RSU}{Resíduos Sólidos Urbanos}
\acrodef{IBGE}{Instituto Brasileiro de Geografia e Estatistica}
\acrodef{IoT}{\textit{Internet of Things}}
\acrodef{LBSNs}{\textit{Location-Based Social Networks}}
\acrodef{LBSN}{\textit{Location-Based Social Network}}
\acrodef{PNRS}{Política Nacional de Resíduos Sólidos}
\acrodef{PEVs}{Pontos de Entrega Voluntária}
\acrodef{PEV}{Ponto de Entrega Voluntária}
\acrodef{SUMO}{\textit{Simulation of Urban MObility}}
\acrodef{PCV}{Problema do Caixeiro Viajante}
\acrodef{PRV}{Problema de Roteamento de Veículos}
\acrodef{AG}{Algoritmos Genéticos}
\acrodef{SPDP}{\textit{Selective Pickup and Delivery Problem}}
\acrodef{MVSPDP}{\textit{Multi-Vehicle Selective Pickup and Delivery Problem}}
\acrodef{FDA}{Função de Distribuição Acumulada Empírica}
\acrodef{CVRP}{Capacitated Vehicle Routing Problem}
\acrodef{PDP}{\textit{Pickup and Delivery Problem}}
\acrodef{G-VRP}{\textit{Green Vehicle Routing Problem}}
\acrodef{AFV}{\textit{Alternative Fuel Vehicles}}
\acrodef{MSLS}{\textit{Multi-Start Local Search}}
\acrodef{AMR}{\textit{Ant Multiple Rounds}}
\acrodef{ACR}{\textit{Ant Colony System}}
\acrodef{SIG}{Sistema de Informação Geográfica}
\acrodef{PSO}{\textit{Particle Swarm Optimization}}
\acrodef{BT}{Busca Tabu}
\acrodef{VNS}{\textit{Variable Neighbourhood Search}}
\acrodef{BFS}{\textit{Best First Search}}
\acrodef{TraCI}{\textit{Traffic Control Interface}}
\acrodef{GDAL}{\textit{Geospatial Data Abstraction Library}}
\acrodef{PMD}{\textit{Política de Menor Distância}}
\acrodef{PMI}{\textit{Política de Menor Impedância}}
\acrodef{PMT}{\textit{Política de Menor Trabalho}}
\acrodef{SPFA}{\textit{Shortest Path Faster Algorithm}}

\newcommand{\trab}{\ac{PMT}\xspace}
\newcommand{\dist}{\ac{PMD}\xspace}
\newcommand{\imp}{\ac{PMI}\xspace}

\title{Sugestões de Rotas Personalizadas para Carrinheiros na Coleta Seletiva de Materiais Recicláveis}

\author{Maria Vitória R. Oliveira\inst{1}, Islene C. Garcia\inst{1}}

\address{Instituto de Computação -- Universidade Estadual de Campinas
  (UNICAMP)\\
  Campinas -- SP -- Brasil
 \email{maria.oliveira@students.ic.unicamp.br, islene@unicamp.br}
}

\begin{document} 

\maketitle

\begin{abstract}
    Carrinheiros are collectors of recyclable materials that use human-powered vehicles. Carrinheiro's collection routes can be tiring depending on the paths chosen. Therefore, this work proposes an algorithm for suggesting customizable routes based on three edge costing policies: Less Work Policy, Less Impedance Policy, and Short Distance Policy. This work used the tools osmnx and networkx to construct graphs, geographic data from Open Street Map, and elevations from Topodata. The simulations performed in \acl{SUMO} (SUMO) demonstrated that the proposed algorithm could minimize the power applied to push the vehicle, the distance, and the travel time, according to the policy used.
\end{abstract}
     
\begin{resumo} 
  Os carrinheiros são catadores de materiais recicláveis que utilizam veículos de tração humana. As rotas de coleta dos carrinheiros podem ser cansativas dependendo do trajeto. Sendo assim, este trabalho propõe um algoritmo de sugestão de rotas personalizáveis a partir de três políticas de cálculo de custos de arestas: Política de Menor Trabalho, Política de Menor Impedância e Política de Menor Distância. Utilizou-se as ferramentas osmnx e networkx na construção dos grafos, dados geográficos do \textit{Open Street Map} e elevações do Topodata. As simulações realizadas no \acl{SUMO} (SUMO) demonstraram que o algoritmo proposto pode minimizar a potência aplicada no veículo, a distância e o tempo de percurso, de acordo com a política utilizada.
\end{resumo}

\section{Introdução} \label{sec:firstpage}

A \ac{PNRS} regulamentou a gestão de \ac{RSU} no Brasil em 2010, estabelecendo diretrizes para a destinação adequada de materiais. De acordo com a \ac{PNRS}, a gestão integrada de \ac{RSU} é um conjunto de ações com o objetivo de encontrar soluções para a destinação dos resíduos, considerando as dimensões política, econômica, ambiental, cultural e social. Além disso, a política determina a inclusão dos catadores de materiais reutilizáveis e recicláveis como agentes no processo de desenvolvimento sustentável. Estes profissionais realizam a coleta seletiva, definida como o processo de recolher os resíduos previamente segredados conforme sua constituição ou composição~\cite{2010lei}.

Os catadores que utilizam veículos de propulsão humana durante o recolhimento dos materiais geralmente são chamados de carrinheiros. Nesse caso, o roteiro é realizado a pé, empurrando o veículo. Sendo assim, a rota de coleta pode ser extremamente cansativa e/ou demorada para o carrinheiro. Portanto, abordar o problema de roteamento de veículos de propulsão humana é importante para proporcionar rotas de coleta eficientes que minimizam o esforço físico ao longo do percurso.

O problema de roteamento de veículos de propulsão humana se divide em dois subproblemas: o primeiro está relacionado à ordenação dos pontos de coleta que devem ser visitados; o segundo está associado à criação de caminhos otimizados entre estes pontos, no mapa geográfico. Na categoria de trabalhos relacionados a ordenação e criação de rotas otimizadas, destaca-se~\cite{vu2018parameter}. Nesse trabalho, apresentou-se um modelo de distribuição geográfica dos pontos de coleta no mapa, maximizando os materiais coletados. Apesar de propor minimizar as rotas dos veículos de tração humana, o trabalho limitou-se a identificar configurações que proporcionam os menores custos financeiros.

Este trabalho propõe um algoritmo de sugestão de rotas para carrinheiros que determina a ordem de visita dos pontos de coleta e define os melhores caminhos entre estes pontos, no mapa geográfico. Além disso, propõe-se personalizar os roteiros de coleta de acordo com três políticas: \trab, \imp e \dist. A política \trab minimiza o trabalho aplicado para empurrar o veículo ao longo do trajeto, a política \imp evita vias que possuem aclives acentuados e a política \dist diminui a distância do roteiro.

As rotas de coleta podem ser personalizadas conforme o objetivo dos carrinheiros e o cenário geográfico, configurando uma das políticas propostas. Por exemplo, se o local de coleta possui declividades acentuadas e/ou o objetivo do catador é diminuir o esforço físico empregado no trajeto, propõe-se utilizar a política \trab ou a \imp. Porém, se o cenário possui baixa variação de altitude e o objetivo do carrinheiro é diminuir a distância percorrida, propõe-se utilizar a política \dist.

O diferencial do algoritmo proposto está na combinação da heurística do Vizinho mais Próximo para determinar a ordem dos pontos de coleta e do algoritmo \ac{SPFA}, para definir os caminhos no mapa geográfico, considerando a utilização de veículos de tração humana. Outro diferencial é a estratégia de atualizar o peso do veículo em cada ponto de parada para identificar os melhores caminhos de acordo com o peso do carrinho e as declividades do cenário, empregada pela política \trab.

Os cenários de avaliação do algoritmo proposto foram implementados com as ferramentas livres de manipulação de grafos networkx e osmnx. Ademais, os dados geográficos foram obtidos do \textit{Open Street Map} e a altitude, do Banco de Dados Geomorfométricos do Brasil (Topodata)~\cite{de2012topodata}. Sendo assim, as políticas \trab, \imp e \dist foram analisadas de acordo com a potência aplicada no veículo, distância e tempo de trajeto. Como resultado, o algoritmo proposto gerou rotas que minimizam a distância e o tempo de trajeto utilizando a política \dist. Porém, a potência média aplicada no veículo foi a maior, em comparação com a potência obtida nas rotas geradas quando as políticas \trab e \imp são empregadas. Além disso, quando a política \trab é implementada, o algoritmo gera rotas que minimizam a potência e a distância.

O restante deste documento está organizado da seguinte forma: a Seção~\ref{sec:trab} apresenta os trabalhos relacionados. Em seguida, a Seção~\ref{sec:estrategias} exibe as estratégias e tecnologias na coleta de resíduos. A Seção~\ref{sec:alg} exibe o algoritmo e as políticas implementadas. Depois, a Seção~\ref{sec:results} apresenta os resultados. Por fim, a Seção~\ref{sec:conclusion} descreve as conclusões do trabalho.

\section{Trabalhos relacionados}
\label{sec:trab}

Diversos trabalhos de otimização no contexto da coleta seletiva de materiais recicláveis se concentram no problema de ordenação e otimização da distribuição geográfica dos pontos de coleta~\cite{rathore2020location, gallardo2015methodology, boskovic2015fast}. Além do problema de distribuição dos pontos de parada, o trabalho~\cite{vu2018parameter} propõe um modelo de roteamento de veículos de tração humana e automotores, para minimizar custos financeiros na coleta de resíduos. O modelo proposto apresentou bom desempenho na distribuição geográfica dos pontos de coleta. Porém, não identificou as rotas que proporcionam as menores distâncias para veículos de tração humana.

Dentre os trabalhos que abordam a definição de rotas para minimizar distâncias nas rotas de coleta seletiva, destaca-se~\cite{Ahmad2020}, que propõe um sistema ideal de recomendação de roteiros para diminuir a distância percorrida, o consumo de combustível do caminhão e maximizar a quantidade de materiais recolhidos. É possível citar também o trabalho~\cite{Benjamin2010}, o qual propõe uma solução considerando janelas de tempo, período de descanso dos condutores e múltiplas instalações de disposição de resíduos. Por fim, o trabalho~\cite{Tavares2009} propõe um modelo para otimização de rotas, minimizando o consumo de combustível dos veículos que coletam e transportam materiais recicláveis.

Outros trabalhos sugerem algoritmos heurísticos para gerar rotas otimizadas considerando veículos automotores, a fim de diminuir a distância percorrida, minimizar os custos financeiros e o consumo de combustível. Sendo assim, o trabalho~\cite{Liao2010} apresenta uma solução logística adaptada à restrição de carga, na qual se deve coletar mercadorias de alguns vértices e cumprir todas as respectivas entregas. Ademais, o trabalho~\cite{Huang2011} propõe um algoritmo com restrições de carga e distância máxima a ser percorrida por cada veículo de uma frota, minimizando a distância total de todos os veículos.

Os trabalhos relacionados estão, em sua maioria, focados no roteamento para veículos automotores. No entanto, é necessário considerar a utilização de veículos de tração humana no roteamento para coleta seletiva, uma vez que os carrinheiros são fundamentais neste processo, especialmente em países emergentes.

\section{Tecnologias na coleta de resíduos}
\label{sec:estrategias}

\begin{figure}[ht]
\centering
\includegraphics[width=.5\textwidth]{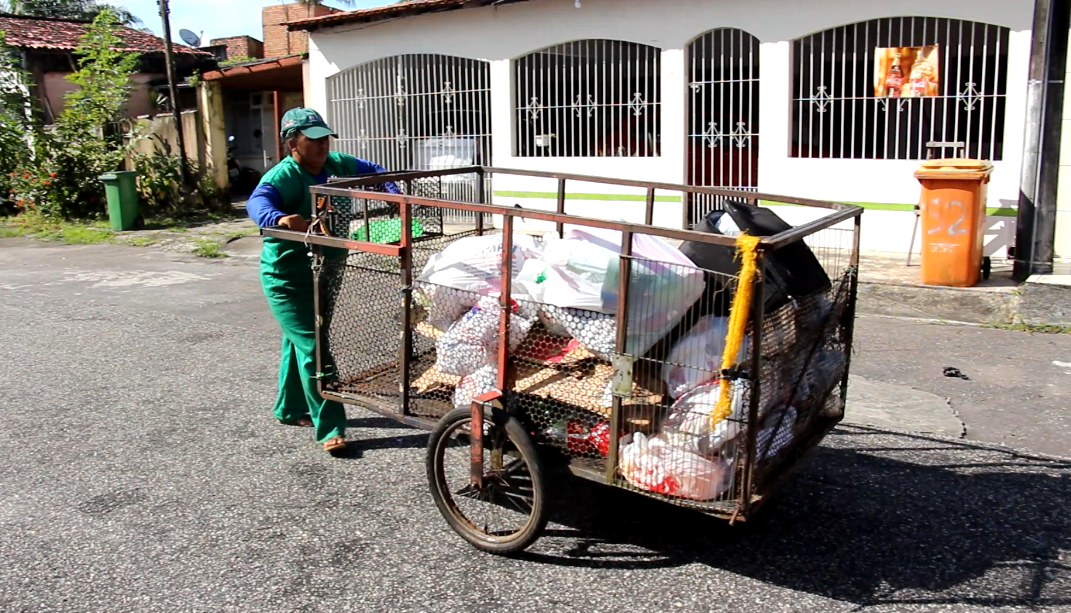}
\caption{Carrinheira utilizando veículo de tração humana na coleta primária}
\label{fig:carrinheira}
\end{figure}

Os carrinheiros geralmente trabalham vinculados a cooperativas ou de forma independente. Nas cooperativas, estão inseridos na metodologia de coleta de três estágios. Esta metodologia pode ser descrita a partir dos processos empregados por uma cooperativa na cidade de Belém, no estado do Pará, Brasil: o caminhão leva os carrinheiros com os veículos de tração humana até um local pré-determinado da cidade. A partir deste local, cada carrinheiro segue uma rota a pé, etapa que também é chamada de coleta primária. A Figura~\ref{fig:carrinheira} apresenta uma carrinheira durante esta fase de coleta. Em seguida, o caminhão transporta os carrinheiros e os resíduos coletados até o galpão de triagem. Após a classificação dos materiais, um veículo de grande porte leva os resíduos até as indústrias de reciclagem.

Na coleta primária, os carrinheiros recolhem os materiais recicláveis na casa das pessoas e estabelecimentos comerciais. Estes locais de parada do roteiro são chamados de pontos de coleta ou pontos de parada. Sendo assim, para que o recolhimento dos materiais aconteça de forma eficiente, o catador precisa saber a localização destes locais previamente. Nesse sentido, existem tecnologias que possibilitam identificar os pontos de coleta, como por exemplo o Cataki~\cite{cataki} e o Destino Sustentável~\cite{wcama}.

Os aplicativos Cataki e Destino Sustentável possibilitam solicitar e agendar coleta de materiais recicláveis com catadores (incluindo carrinheiros). Para isso, o solicitante precisa informar o tipo de resíduo, a quantidade/peso dos materiais, os dias e horários disponíveis para entregar os materiais ao catador, assim como o local de entrega. No Cataki, ao solicitar a coleta, o aplicativo indica um catador geograficamente próximo para recolher os materiais no local cadastrado. Após selecionar o(a) catador(a), a coleta fica registrada na agenda de quem vai coletar e do solicitante. No Destino Sustentável, este processo ocorre de maneira parecida. Porém, após solicitar a coleta, um catador precisa atender o anúncio para que o recolhimento do resíduo seja agendado~\cite{wcama}.

Com a agenda definida, os catadores podem decidir o roteiro de coleta previamente. Para isso, é preciso definir a ordem de visitação dos pontos de coleta considerando o acréscimo de peso ao veículo em cada ponto de parada. Sendo assim, o peso do veículo depende da produtividade. Um carrinheiro vinculado a uma cooperativa considerada de alta eficiência recolhe, em média, 85,7 quilos de resíduos por dia~\cite{silva2017organizaccao}. Nessa perspectiva, muitas vezes os carrinheiros criam roteiros baseados no conhecimento empírico sobre o local.

\section{Algoritmo e políticas implementadas}
\label{sec:alg}

Esta seção descreve o algoritmo proposto e as políticas implementadas. O algoritmo proposto gera rotas otimizadas utilizando uma das políticas (\trab, \imp e \dist). Estas políticas descrevem o mecanismo de atualização de peso das arestas dos grafos. Sendo assim, é possível dividir o algoritmo proposto nas seguintes etapas: cálculo da área que compreende o trajeto; obtenção dos dados geográficos; construção dos grafos; ordenação das visitas aos pontos de coleta e criação da rota.

Dado um conjunto de coordenadas geográficas, a primeira representa o ponto de início e a última, o destino final. As outras coordenadas representam os pontos de coleta. Com base nisso, o algoritmo delimita uma área retangular com os valores mínimos e máximos de latitude e longitude do conjunto de coordenadas. Esta área é utilizada para obtenção de dados geográficos do \textit{Open Street Map} e de altitude, em formato GeoTiff, no Banco de Dados Geomorfométricos, do projeto Topodata~\cite{de2012topodata}. Em seguida, emprega-se a ferramenta osmnx para gerar o grafo da região delimitada, chamado grafo geográfico. Nesse caso, os vértices representam as esquinas e as arestas, as ruas~\cite{boeing}. Depois, identifica-se os valores de altitude de cada vértice do grafo geográfico para configurar os ângulos das arestas.

Os dados geográficos do \textit{Open Street Map} adicionados no grafo são as coordenadas dos vértices, a velocidade máxima das vias, o tipo de pavimentação e o tamanho das arestas. Vias com velocidades máximas de $40$ Km/h ou menos são configuradas como bidirecionais, possibilitando ao carrinheiro andar nos dois sentidos da rua. Além disso, os pontos de parada são adicionados no grafo, sendo que o primeiro ponto é chamado de vértice de início e o último, de vértice depósito (local de disposição dos materiais).

\begin{figure}[ht]
\centering
\subfigure[Grafo para ordenação]{
\includegraphics[width=4.5cm]{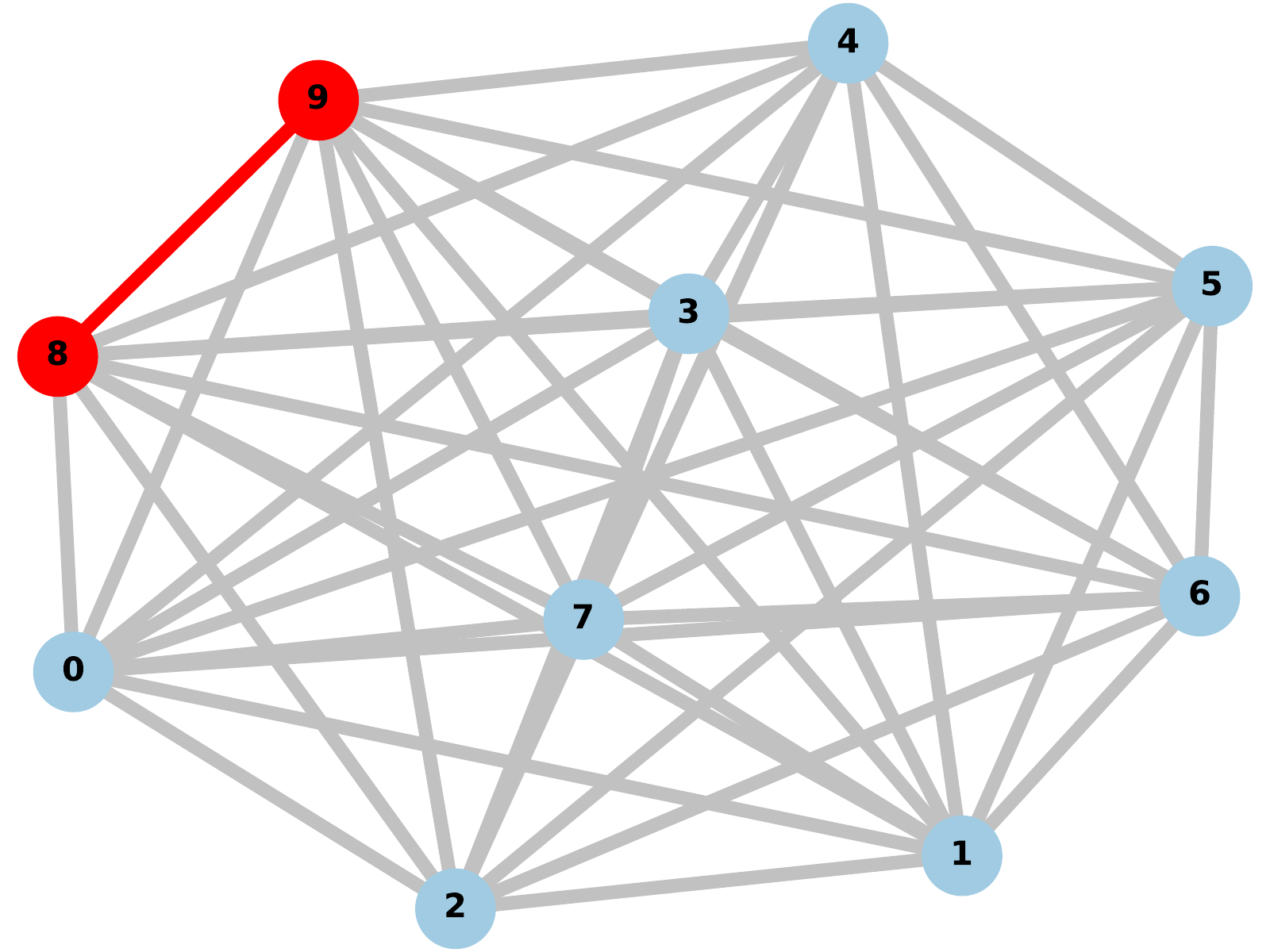}
\label{fig:grafo_ordenacao}
}
\subfigure[Grafo geográfico]{
\includegraphics[width=5cm]{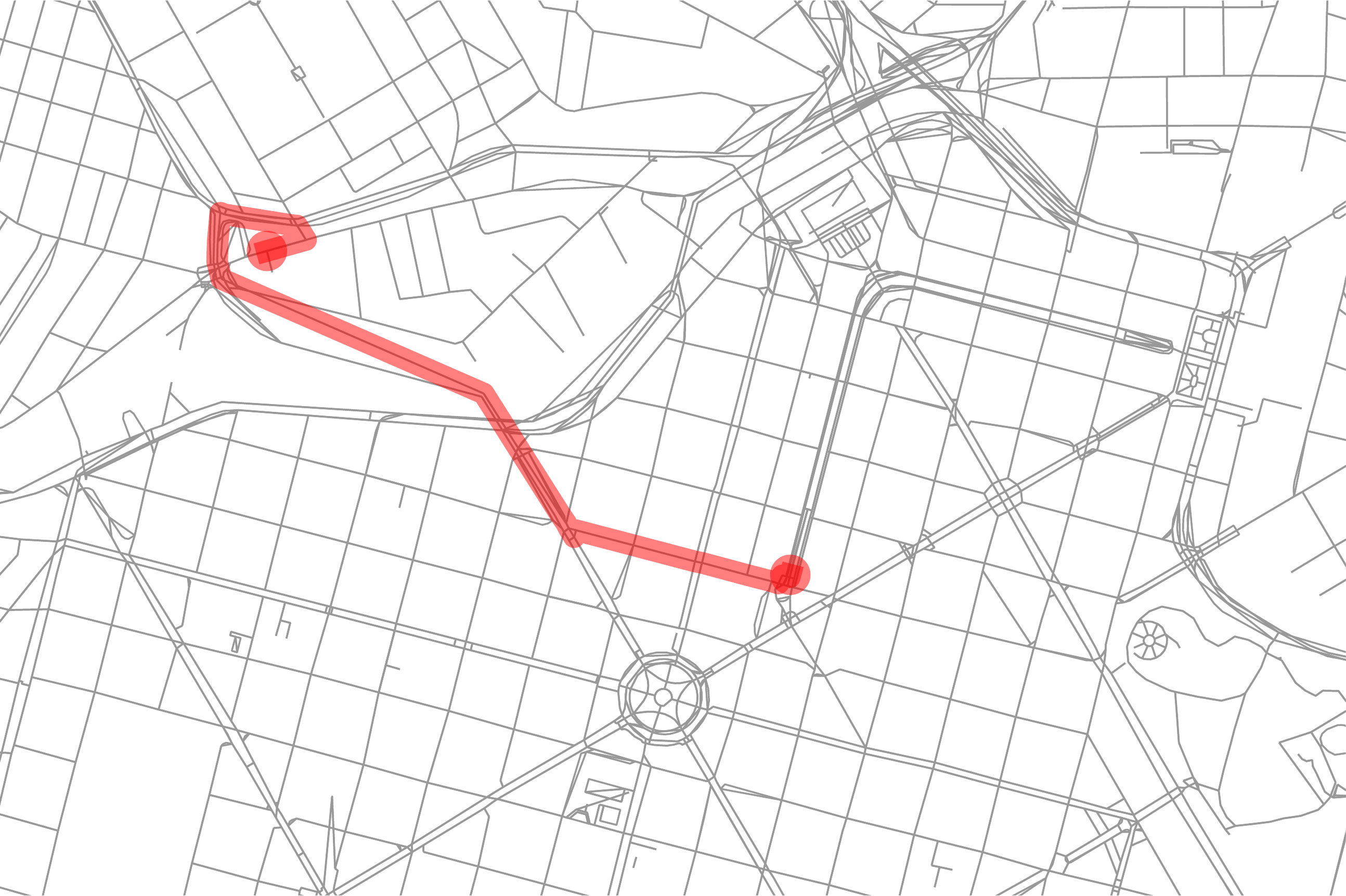}
\label{fig:grafo_cenario}
}

\label{fig:grafos}

\caption{Ilustração dos grafos utilizados para determinação da rota}

\end{figure}

Para definir a ordem de visitação dos pontos de coleta, cria-se um grafo completo utilizando o networkx, nomeado de grafo para ordenação, com o número de vértices igual ao número de pontos de parada mais os pontos inicial e final. Por ser completo, cada vértice é adjacente a todos os outros. Nesse sentido, o custo das arestas do grafo para ordenação é calculado a partir do trajeto entre os pontos de coleta, no grafo geográfico. A Figura~\ref{fig:grafo_ordenacao} ilustra um exemplo de grafo para ordenação, com dois pontos de coleta destacados. O custo da aresta entre os vértices 8 e 9 é calculado de acordo com o custo total do caminho entre os dois pontos de coleta no grafo geográfico. A Figura~\ref{fig:grafo_cenario} apresenta a rota que conecta os vértices 8 e 9 no grafo geográfico destacada em vermelho.

Os caminhos no grafo geográfico são definidos a partir do algoritmo \ac{SPFA}, baseado no algoritmo de Bellman-Ford. Este algoritmo foi escolhido porque aceita pesos negativos de arestas~\cite{moore1959shortest, madkour2017survey}. Além disso, a ordenação dos pontos de coleta é realizada utilizando a heurística do Vizinho Mais Próximo, que é amplamente utilizada em problemas de roteamento de veículos, possui baixo tempo de processamento e bom desempenho~\cite{daanish2017implementation}.

Descreve-se a a definição das rotas nas seguintes etapas: partindo-se do nó inicial, verifica-se o custo desse nó até cada um dos vértices adjacentes que não foram visitados ainda. Esse custo é calculado de acordo com algoritmo \ac{SPFA}, que escolhe o melhor caminho entre o par de vértices, no grafo geográfico. Em seguida, encontra-se o vértice mais próximo do nó atual, ou seja, que possui o menor custo. O nó mais próximo passa a ser o vértice atual e, novamente, calcula-se os custos até os vértices que não foram visitados. Esse processo é realizado até todos que os pontos de coleta sejam inseridos no caminho. Visto que o nó depósito sempre será o último vértice, ele é inserido no final.

O mecanismo de geração de rotas está descrito mais detalhadamente no Algoritmo~\ref{alg:algoritmo}. Além disso, a Tabela \ref{tab:sumario} apresenta um sumário dos símbolos utilizados e seus respectivos significados.

\begin{table}[ht]
\centering
\caption{Sumário dos símbolos e significados utilizados nesta seção}
\label{tab:sumario}
\begin{tabular}{ll}
\textbf{Símbolo}      & \textbf{Significado}                                   \\
Y            & Conjunto de vértices visitados do grafo para ordenação         \\
G            & Grafo geográfico                                             \\
V            & Conjunto de vértices do grafo para ordenação (exceto nó depósito) \\
T            & Conjunto de vértices não visitados do grafo para ordenação (exceto nó depósito) \\ 
$t_{i}$      & Elemento i de T                                              \\
$Y_{j}$      & Último vértice adicionado em Y (vértice atual)                     \\
K            & Caminho geográfico entre dois pontos de parada               \\
$K_{min}$    & Menor caminho entre dois pontos de parada                    \\
$C_{k}$      & Custo do caminho K                                           \\
$C_{min}$    & Custo mínimo até o próximo vértice                          \\
n            & Vértice mais próximo do nó atual ($Y_{j}$)                    \\
início       & Vértice de início                                             \\ 
fim          & Vértice depósito                                             \\    
\end{tabular}
\end{table}

\begin{algorithm}[ht]
\caption{Heurísticas de identificação das rotas}
\label{alg:algoritmo}
\SetAlgoLined
\Entrada{início, fim, G, V, T, política, massaVeiculo}
\Saida{rota}

Y $\gets$ {início} \\
n $\gets \phi $ \\
rota $ \gets \phi $ \\

\Enqto{Y $ \not= $ V}{

    $C_{min}$ $\gets  \infty $ \\
    \Para{cada $t_{i}$ em T}{
         K $ \gets $ shortestPathFaster($Y_{j}$, $t_{i}$, G, política) \\
         \Se{$C_{k}$ $ < $ $C_{min}$}{
            $C_{min}$ $ \gets $ $C_{k}$ \\
            $K_{min}  \gets $ K \\
            n $ \gets $ $t_{i}$ \\ 
         }
    }
    rota $ \gets $ rota $ \cup  K_{min} $ \\
    Y $ \gets $ Y $ \cup $ \{n\}\\
    \Se{política $=$ \trab}{
        massaVeiculo $ \gets $ atualizaMassa(massaVeiculo, $n$) \\
        G $ \gets $ atualizaPesoArestas(G, massaVeiculo) \\
    }
}    
$K_{min} $ $\gets$ shortestPathFaster($Y_{j}$, fim, G, política) \\
rota $ \gets $ rota $ \cup$ $ K_{min} $ \\
\Retorna{rota}
    
\end{algorithm}

O Algoritmo~\ref{alg:algoritmo} descreve detalhes da combinação da heurística Vizinho Mais Próximo com o algoritmo SPFA. Os parâmetros de entrada do Algoritmo~\ref{alg:algoritmo} são: vértice inicial (início), vértice final (vértice depósito), o grafo geográfico ($G$); conjunto de vértices do grafo para ordenação ($V$); conjunto de vértices que ainda não foram visitados do grafo para ordenação ($T$); a política (\trab, \imp ou \dist) e a massa inicial do veículo.

De acordo com o Algoritmo~\ref{alg:algoritmo}, enquanto o conjunto de vértices visitados ($Y$) for diferente do conjunto de vértices do grafo para ordenação ($ V $) (linha 4), configura-se o custo mínimo até o nó mais próximo ($C_{min}$) igual a infinito (linha 5). Para cada vértice não visitado $t_{i}$ (linha 6), identifica-se o caminho ($K$) entre o nó atual ($Y_{j}$) e $t_{i}$. Este caminho é calculado por meio do algoritmo \ac{SPFA} (linha 7). Se o custo do caminho ($C$) for menor que o custo mínimo ($C_{min}$), atualiza-se o custo mínimo ($C_{min}$) (linha 8), o menor caminho ($K_{min}$) e o nó mais próximo ($n$) (linhas 10 e 11).

Depois de identificar o vértice mais próximo ($n$) do nó atual ($Y_{j}$), adiciona-se o menor caminho ($K_{min}$) na rota (linha 14). Em seguida, o nó mais próximo ($n$) é acrescentado ao conjunto de vértices visitados ($Y$) (linha 15), tornando-se o vértice atual ($Y_{j}$). Além disso, se a política utilizada for a \trab, a massa do veículo é atualizada de acordo com a função $atualizaMassa$. Esta função verifica qual o valor de massa dos materiais que serão coletados no ponto de coleta $n$ e acrescenta este valor à massa do veículo atual (linha 17). Ademais, atualiza-se os pesos das arestas do grafo geográfico a partir do novo valor de massa do veículo (linha 18). Quando todos os vértices contidos em $V$ forem visitados, cria-se o caminho do último nó adicionado ($Y_{j}$) até o nó depósito (linha 21). Acrescenta-se este caminho na rota (linha 22) e, por fim, o algoritmo retorna a rota. O código completo do algoritmo está disponível no GitHub~\footnote{https://github.com/vivirodrigues/Carrinheiros}.

O Algoritmo~\ref{alg:algoritmo} apresentou o processo de geração de rotas utilizando a heurística do Vizinho Mais Próximo e o algoritmo \ac{SPFA}. Além disso, foi possível identificar que a política \trab atualiza o peso do veículo e das arestas cada vez que um nó é visitado no grafo para ordenação.

\subsection{Política de Menor Trabalho}

A política \trab calcula o peso das arestas do grafo geográfico a partir do trabalho necessário para empurrar o veículo em determinada aresta. De acordo com a mecânica newtoniana, este cálculo pode ser realizado por meio da equação~\ref{eq: trab}.

\begin{equation}
\label{eq: trab}
    W = \left[(m * g * f_{r} \cos \theta) + (m * g * \sin \theta) + \left(\frac{1}{2} * \rho * C * S * v^{2}\right)\right] * d
\end{equation}

Na qual $W$ representa o valor do Trabalho, $m$ o valor da massa do veículo, $g$ a constante de aceleração da gravidade, $f_{r}$ o coeficiente de resistência ao rolamento, $ \theta $ o ângulo de inclinação da via, $ \rho $ a densidade do ar, $C$ corresponde ao coeficiente de arrasto, $S$ a área frontal do veículo, $v$ a velocidade do ar em relação ao corpo e $d$ representa a distância total da aresta~\cite{silveira2011potencia}~\cite{halliday2016fundamentos}.

A primeira componente da equação corresponde à força de resistência ao rolamento, associada à perda de energia mecânica na roda, que ocorre em razão da deformação do pneu em contato com o solo. Nesse caso, o valor de $f_{r}$ depende do tipo de roda do veículo e da pavimentação. Além disso, a segunda componente representa a força de ação da gravidade, a qual tem influência resistente em aclives e adjuvante em declives. Na terceira componente, tem-se a força de resistência do ar em relação ao veículo em movimento, chamada de força de arrasto. Considera-se que a velocidade do ar em relação ao corpo ($v$) é igual a velocidade do carrinho, consequentemente, o ar está parado em relação à via~\cite{silveira2011potencia}.

\subsection{Política de Menor Impedância}

A política \imp evita aclives acentudados e foi idealizada para ser utilizada quando os carrinheiros possuem limitações físicas ou preferem caminhos com poucas subidas. Nesse sentido, propõe-se calcular o valor da Impedância ($ I $) de acordo com a equação~\ref{eq:imped}.

\begin{equation}
    I = 
\begin{cases}
\label{eq:imped}
    \theta^2 * d, & \text{se } \theta > 0\\
    \theta * (-1) * d, & \text{caso contrário}
\end{cases}
\end{equation}

Na qual $ \theta $ representa o ângulo da via e $ d $, o tamanho da aresta (distância). A partir da equação~\ref{eq:imped}, tem-se um valor constante de peso das arestas, pois não precisa ser atualizado durante o percurso, diferentemente do que ocorre com a política \trab.

\subsection{Política de Menor Distância}

A política \dist minimiza a distância percorrida. Neste caso, o peso das arestas é igual ao valor da distância/tamanho da aresta, obtida por meio do \textit{Open Street Map}.

\section{Resultados}\label{sec:results}

Esta Seção apresenta os parâmetros utilizados nas simulações e os resultados obtidos.

\subsection{Parâmetros das simulações}

\begin{figure}[ht]
\centering
\subfigure[Cenário Belo Horizonte]{ % 7.2
\includegraphics[width=6.8cm]{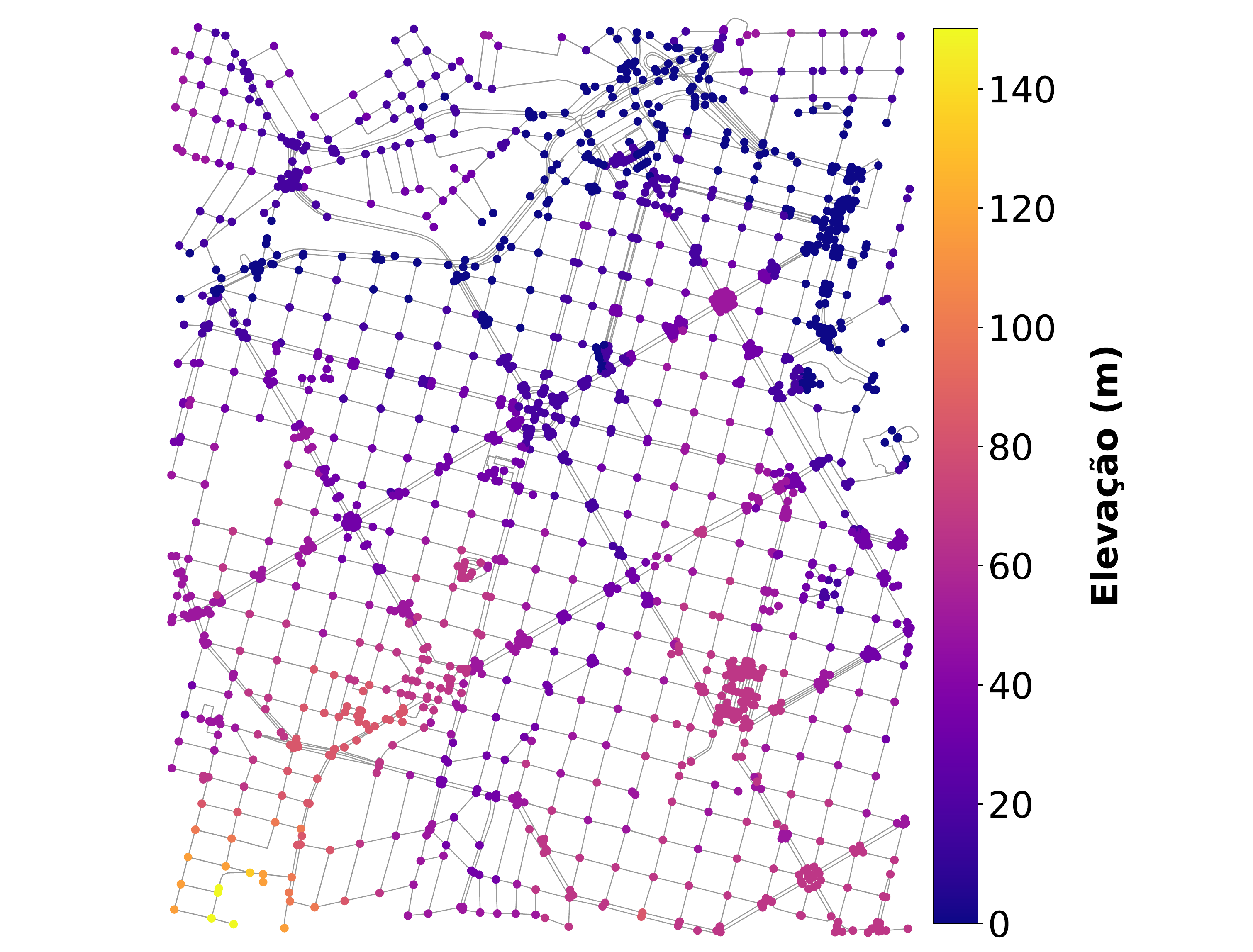}
\label{fig:subf_a}
}
\subfigure[Cenário Belém]{
\includegraphics[width=6.8cm]{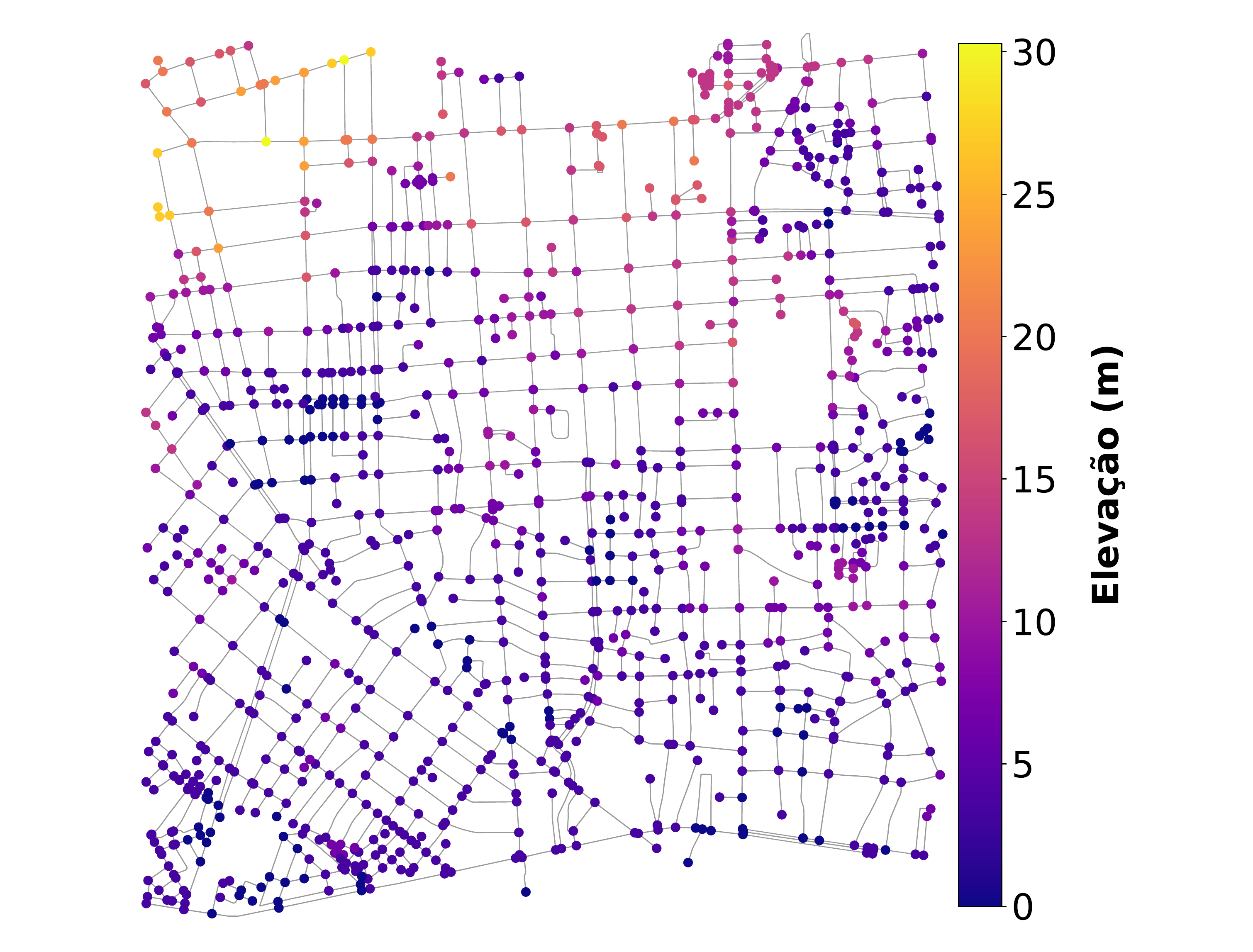}
\label{fig:subf_b}
}

\caption{Cenários de simulação}
\end{figure}

Dois cenários foram utilizados para validar a proposta, com diferentes variações de altitude, ambos no Brasil. O primeiro é na cidade de Belo Horizonte, no estado de Minas Gerais, e o segundo é na cidade de Belém, no estado do Pará. Ilustra-se os grafos dos cenários de Belo Horizonte na Figura~\ref{fig:subf_a} e Belém na Figura~\ref{fig:subf_b}, a partir dos dados de altitude normalizada. De acordo com a Figura~\ref{fig:subf_a}, é possível perceber que a elevação varia de 0 a 150 metros no cenário de Belo Horizonte. Na Figura~\ref{fig:subf_b}, que representa o cenário de Belém, a elevação varia de 0 a 30 metros. Nesse sentido, o cenário de Belo Horizonte possui maior variação na altitude dos nós.

As rotas foram geradas pelo algoritmo proposto e validadas no simulador \ac{SUMO}. De acordo com a Tabela~\ref{tab:simulacao}, a qual demonstra os parâmetros utilizados nas simulações, utilizou-se 8 pontos de coleta em cada simulação. Nesses pontos, a massa do veículo é incrementada de forma pseudo-aleatória, com valor máximo de $50$ Kg. Este valor de atualização da massa do veículo representa os materiais recicláveis coletados. Além dos 8 pontos de coleta, define-se também o ponto inicial e final da rota. As coordenadas $x$ e $y$ desses pontos seguem uma distribuição gaussiana com desvio padrão $ 0.005 $, média ($-43.9438, -19.9202$), no cenário de Belo Horizonte, e média ($-48.47000, -1.46000$), no cenário de Belém. Ademais, a massa inicial do veículo foi configurada como $110$ Kg.

Para avaliar os resultados, escolheu-se as métricas: potência aplicada para empurrar o veículo, distância total do trajeto e tempo de percurso. O cálculo da potência instantânea foi realizado a partir da equação~\ref{eq: power}~\cite{silveira2011potencia} e implementado na simulação por meio da ferramenta Traci do simulador \ac{SUMO}. As simulações foram executadas 30 vezes utilizando diferentes sementes de aleatoriedade e intervalo de confiança de 95\%.

\begin{equation}
\label{eq: power}
    P = \left[(m * g * f_{r} \cos \theta) + (m * g * \sin \theta) + \left(\frac{1}{2} * \rho * C * S * v^{2}\right)\right] * v
\end{equation}

Na qual $P$ representa a potência instantânea, $m$ o valor da massa do veículo, $g$ a constante de aceleração da gravidade, $f_{r}$ o coeficiente de resistência ao rolamento, $ \theta $ o ângulo de inclinação da via, $ \rho $ a densidade do ar, $C$ corresponde ao coeficiente de arrasto, $S$ a área frontal do veículo e $v$ a velocidade instantânea do veículo, que é definida automaticamente pelo algoritmo do simulador SUMO. Como mostra a Tabela~\ref{tab:simulacao}, configurou-se a velocidade máxima do veículo como $3.6$ Km/h, a área frontal do veículo $ 1 m^{2}$, a densidade do ar $ 1.2 Km/m^{3} $ e o coeficiente aerodinâmico $1$. Por fim, o coeficiente de resistência ao rolamento foi definido de acordo com o tipo de pavimentação da via~\cite{costa2015proposta}.

\begin{table}[ht]
\centering
\caption{Parâmetros utilizados nas simulações}
\label{tab:simulacao}
\begin{tabular}{ll}
\textbf{Parâmetro}      & \textbf{Valor}                                   \\
Número de pontos de coleta por simulação     & 8  \\
Incremento máximo de massa do veículo em cada ponto de coleta  & 50 Kg \\
Média das coordenadas x e y Belo Horizonte   & (-43.9438, -19.9202)   \\
Média das coordenadas x e y Belém      & (-48.47000, -1.46000) \\
Desvio padrão das coordenadas    & 0.005   \\
Velocidade máxima do carrinheiro            & 3.6 Km/h         \\
Peso inicial do veículo             & 110 Kg \\
Área frontal do veículo            & 1 m² \\
Densidade do ar           & 1.2 Km/m³\\
Coeficiente aerodinâmico   & 1 \\
Aceleração da gravidade & 9.80665 \\

\end{tabular}
\end{table}

\subsection{Potência aplicada no veículo}

\begin{figure}[ht]
\centering
\includegraphics[width=.6\textwidth]{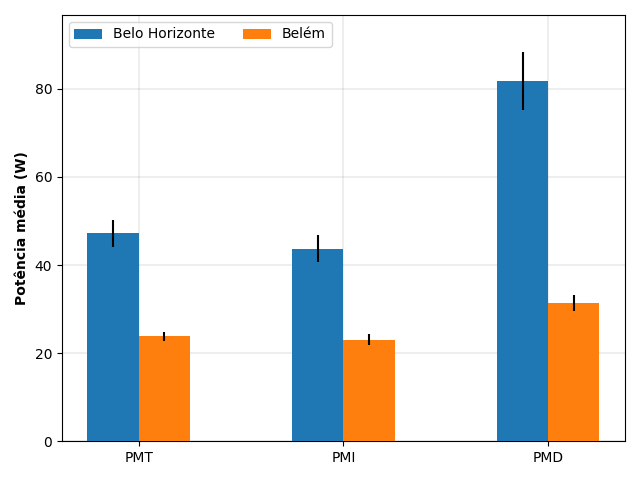}
\caption{Potência média empregada para empurrar o veículo}
\label{fig:final}
\end{figure}

A potência aplicada no veículo depende da força de tração e da velocidade empregada pelo carrinheiro. Visto que é necessário aplicar esforço físico para locomover os veículos, os trajetos de coleta tornam-se mais cansativos quando se  emprega altas potências. Sendo assim, a Figura~\ref{fig:final} apresenta os resultados da potência média empregada no veículo conforme as simulações das rotas geradas pelo algoritmo proposto, utilizando as políticas \trab, \imp e \dist.

De acordo com a Figura~\ref{fig:final}, os valores de potência média aplicada no veículo nas rotas geradas utilizando a política \trab nos cenários de Belo Horizonte e Belém foram 47.1 W e 23.8 W, respectivamente. Considerando as rotas geradas pelo algoritmo com a política \imp, a potência média aplicada foi de 43.7 W no cenário de Belo Horizonte e 23.09 W, no cenário de Belém. Os valores de potência média aplicada no veículo nas rotas geradas pelo algoritmo utilizando a política \dist foram 81.7 W e 31.4 W, nos cenários de Belo Horizonte e Belém, respectivamente.

Os resultados da Figura~\ref{fig:final} demonstram que os valores de potência média aplicada no veículo nas rotas do cenário com menor variação de altitude (Belém) foram menores do que no cenário com maior variação de altitude (Belo Horizonte). Verifica-se, então, que os aclives e declives influenciam na potência média aplicada no veículo.
O algoritmo gerou rotas com valores semelhantes de potência média aplicada no veículo utilizando as políticas \trab e \imp, tanto no cenário de Belo Horizonte, quanto Belém. Sendo assim, considerar a declividade do local na escolha do trajeto diminuiu a potência média empregada para empurrar o veículo. Além disso, as rotas geradas pelo algoritmo utilizando a política \imp obtiveram os maiores valores de potência média nos dois cenários. Portanto, as rotas que priorizaram a menor distância de trajeto exigiram maior esforço físico nos cenários implementados.

\begin{figure}[ht]
\centering
\subfigure[Belo Horizonte]{
\includegraphics[width=7.1cm]{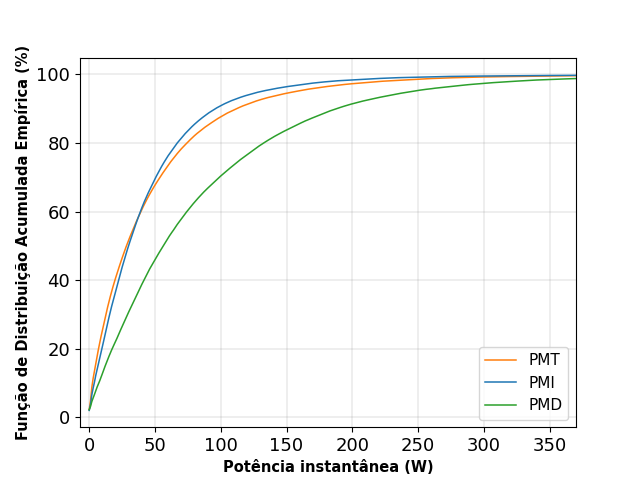}
\label{fig:cdf_bh}
}
\subfigure[Belém]{
\includegraphics[width=7.1cm]{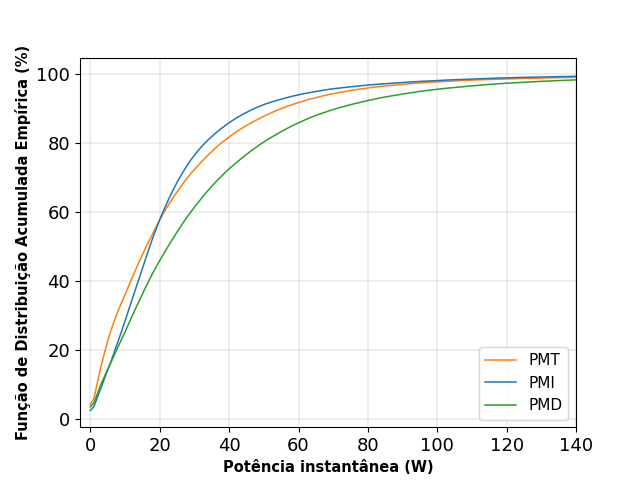}
\label{fig:cdf_belem}
}
\label{fig:cdf}

\caption{Função de Distribuição Acumulada Empírica da potência instantânea aplicada}

\end{figure}

Além da potência média, é preciso analisar os valores de potência instantânea aplicada no veículo durante as simulações. Para isso, a Figura~\ref{fig:cdf_bh} e a Figura~\ref{fig:cdf_belem} apresentam os resultados da Função de Distribuição Acumulada (FDA) Empírica da potência instantânea aplicada no veículo nos cenários de Belo Horizonte e Belém.

De acordo com os resultados da Figura~\ref{fig:cdf_bh}, as rotas geradas pelo algoritmo utilizando a política \imp proporcionaram 69\% de probabilidade de aplicar valores menores ou iguais a 50 W de potência instantânea no veículo e 90\% de probabilidade de aplicar potências instantâneas menores ou iguais a 100 W.
As rotas geradas pelo algoritmo utilizando a política \trab apresentaram resultados parecidos com as rotas da política \imp: 67\% de probabilidade do valor de potência aplicada no veículo ser menor ou igual a 50 W e 87\% de probabilidade de ser menor ou igual a 100 W.
Além disso, as rotas geradas pela política \dist obtiveram 45\% e 70\% de probabilidade de aplicarem potências instantâneas menores ou iguais a 50 W e 100 W, respectivamente.

Observa-se na Figura~\ref{fig:cdf_bh} que a FDA empírica se aproxima de 100\% a partir do valor 250 W nas políticas \trab e \imp, enquanto que na política \dist, a FPA empírica se aproxima de 100\% após 350 W. Além disso, os resultados das simulações utilizando a política \imp demonstraram menor probabilidade do veículo apresentar potências instantâneas mais baixas. Considera-se, então, que as rotas dessa configuração exigem maior esforço físico do carrinheiro para empurrar o veículo.

Os resultados da Figura~\ref{fig:cdf_belem} demonstram que a política \trab proporcionou a obtenção de rotas com 87\% de probabilidade de empregar 50 W ou menos de potência instantânea no veículo e 97\% de probabilidade de aplicar potências menores ou iguais a 100 W.
Além disso, as rotas geradas com a política \imp apresentaram 91\% de probabilidade de aplicar 50 W ou menos de potência instantânea no veículo e 98\% de probabilidade de aplicar valores menores ou iguais a 100 W.
Por fim, as rotas geradas pelo algoritmo adotando a política \dist apresentaram 80\% de probabilidade de empregar valores menores ou iguais a 50 W de potência instantânea no veículo e 95\% de probabilidade de aplicar potências instantâneas menores ou iguais a 100 W.

Em conformidade com os resultados de potência média apresentados na Figura~\ref{fig:final}, as rotas geradas utilizando as políticas \trab e \imp no cenário com menor variação de altitude alcançaram resultados aproximados.
Além disso, a variação de potência instantânea aplicada no cenário Belém é de 0 a aproximadamente 140~W. Portanto, há menor variação de potência aplicada neste cenário, em comparação com o cenário de Belo Horizonte, que é aproximadamente 350~W.

\subsection{Tempo de trajeto e distância total do roteiro}

\begin{figure}[h!]
\centering
\subfigure[Distância]{
\includegraphics[width=7cm]{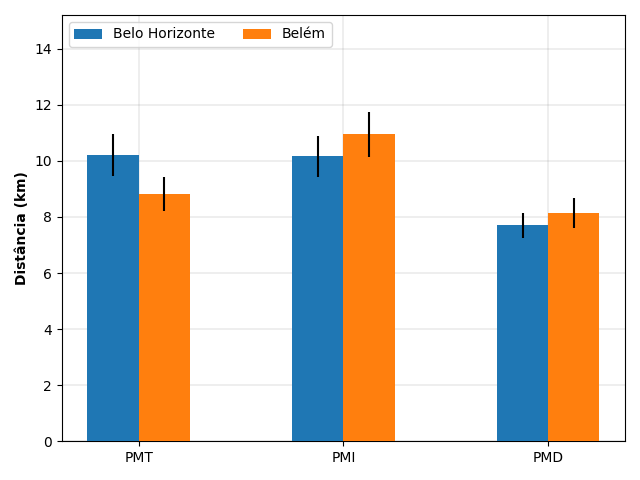}
\label{fig:distancia}
}
\subfigure[Tempo]{
\includegraphics[width=7cm]{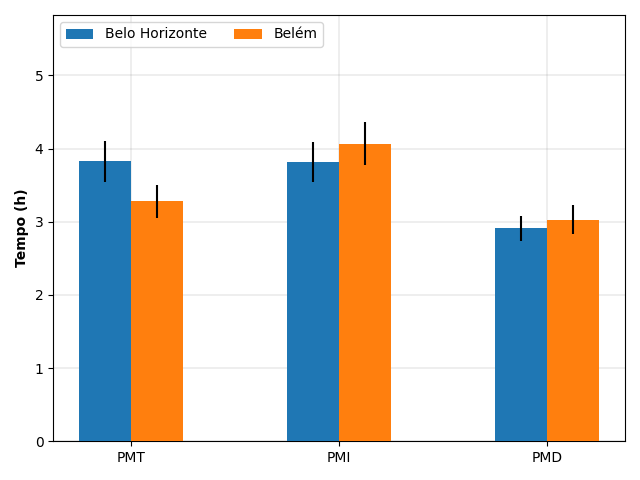}
\label{fig:tempo}
}
\label{fig:a}

\caption{Distância e Tempo de trajeto}

\end{figure}

A avaliação da distância e tempo total do trajeto permitem identificar a política mais adequada para os carrinheiros que preferem rotas mais curtas e rápidas. Com base nos resultados das simulações, a Figura~\ref{fig:distancia} apresenta os resultados da distância total dos roteiros e a Figura~\ref{fig:tempo} demonstra os resultados de tempo de trajeto.

De acordo com a Figura~\ref{fig:distancia}, a média da distância percorrida pelo carrinheiro na simulação das rotas geradas pelo algoritmo utilizando a política \trab nos cenários de Belo Horizonte e Belém foram 10.2 Km e 8.8 Km, respectivamente. Além disso, a política \imp proporcionou rotas com média de distância total de 10.16 Km no cenário de Belo Horizonte e 10.9 Km no cenário de Belém. As rotas geradas empregando a política \dist alcançaram 7.7 Km e 8.1 Km de distância. Em seguida, a Figura~\ref{fig:tempo} apresenta a comparação entre as médias de tempo total de trajeto das rotas geradas pelo algoritmo aplicando as três políticas propostas. As médias do tempo de percurso das rotas geradas utilizando a política \trab foram 3.8 horas e 3.2 horas, nos cenários de Belo Horizonte e Belém, respectivamente. A política \imp proporcionou rotas com média de 3.8 horas no cenário de Belo Horizonte e 4.06 horas no cenário de Belém. Por fim, o algoritmo utilizando a política \dist gerou rotas com médias 2.9 e 3.02 nos cenários de Belo Horizonte e Belém, respectivamente.

As rotas geradas utilizando a política \dist obtiveram a menor média de tempo e distância nos dois cenários. Visto que o resultado das políticas \trab e \imp foram muito próximas no cenário de Belo Horizonte, observa-se que para evitar vias que demandam maior esforço físico, como aclives acentuados, estas políticas proporcionaram rotas um pouco mais longas. No entanto, no cenário de Belém, o tempo de trajeto e distância percorrida das rotas geradas pela política \trab foram menores comparadas com as rotas geradas pela política \imp. Portanto, no cenário de menor variação de altitude (Belém), além da política \trab proporcionar rotas que minimizam eficientemente a potência média, alcançou também rotas mais rápidas e menores do que a política \imp. Sendo assim, a política \trab é mais eficiente para minimizar o esforço físico do carrinheiro.

\section{Conclusão}\label{sec:conclusion}

Este trabalho apresentou uma abordagem original de sugestão de rotas para catadores que utilizam veículos de propulsão humana na coleta seletiva, desenvolvida a partir das tecnologias livres networkx, osmnx e \textit{Open Street Map}. O algoritmo proposto gera rotas de coleta utilizando as heurísticas do Vizinho mais Próximo para ordenar a visita aos pontos de coleta e o algoritmo \ac{SPFA}, para determinar o melhor caminho entre pontos de coleta no mapa geográfico. Além disso, três políticas de personalização de rotas foram propostas: Política de Minimização do Trabalho, Política de Minimização da Impedância e Política de Minimização da Distância.

A política \trab obteve o melhor desempenho quanto à minimização do esforço físico porque, além de apresentar bom desempenho na diminuição da potência aplicada no veículo, proporcionou rotas com menor distância e tempo, em comparação com a política \imp no cenário de Belém. Além disso, as rotas mais curtas e rápidas foram obtidas a partir da política \dist. No entanto, estas rotas apresentaram os maiores valores de potência aplicada. Demonstrando, assim, que os caminhos de menor distância não foram os que diminuíram o esforço físico do carrinheiro nos cenários considerados.

Este trabalho apresenta relevância social, pois possibilita melhorar a qualidade de vida dos carrinheiros por meio de sugestões de rotas que minimizam o esforço físico. Nessa perspectiva, como trabalhos futuros, propõe-se sugestões de rotas que evitam vias de tráfego em alta velocidade, para minimizar os riscos de acidentes.

\bibliographystyle{sbc}
\bibliography{sbrc_bib}

\end{document}